\documentstyle[12pt,a4]{article}
\begin{document}
\def\doublespaced{\baselineskip=\normalbaselineskip\multiply\baselineskip
  by 150\divide\baselineskip by 100}
\doublespaced
\def\lsim{~{\rlap{\lower 3.5pt\hbox{$\mathchar\sim$}}\raise 1pt\hbox{$<$}}\,}
\def\gsim{~{\rlap{\lower 3.5pt\hbox{$\mathchar\sim$}}\raise 1pt\hbox{$>$}}\,}
\def\thisday{~\today ~and~ hep-pth/yymmnnn~~}

\begin{titlepage}
\vspace{0.5cm}
\begin{center}
TREATING 'tHOOFT-POLYAKOV MONOPOLE AS CONSTRAINED SYSTEM \\
Hatem Widyan\footnote{email: widyan@ahu.edu.jo} \\
{{Department of Physics \\
Al-Hussein Bin Talal University \\
P.O. Box 20 \\
 Ma'an, Jordan}}\\
\end{center}
\vspace{0.4cm}
\raggedbottom
\relax

\begin{abstract}
\noindent The 'tHooft-Polyakov monopole is treated as constrained
system using the Hamilton-Jacobi method. The set of the
Hamilton-Jacobi partial differential equations and the equations
of motion are obtained. The quantization of the system is also
discussed.
\end{abstract}

\vspace{0.5cm}

\vspace{1.0cm}
\end{titlepage}

\section*{1 Introduction}

  A physical system whose space consists of $2N$ degrees of freedom:
  $q=(q_i, \cdots,q_N)$, $p=(p_1,\cdots, p_N)$ is called a constrained
  system if there are relations between coordinates and momenta. In other
  words, in constrained systems some of the velocities $\dot{q}_i(i=1,\cdots,
  N)$ cannot be expressed in terms of the coordinates and momenta. When
  this happens, the Lagrangian of the system is called singular. The
  constraints place restrictions on the possible choices of boundary
  conditions for the canonical coordinates. Moreover, the standard
  quantization methods cannot be applied directly to constrained theories.
  However, the basic ideas of the classical treatment and the quantization
  of such systems were presented first by Dirac~\cite{dirac}.
  Following Dirac, it was shown (Faddeev~\cite{fadeev}, Hanson
  {\it et al}~\cite{hanson}, Dayi~\cite{dayi}, Evans~\cite{evans}) that
  gauge fixing conditions should be imposed for first class constraints.

   The construction of the Hamilton-Jacobi partial differential equations
   (HJPDEs)  for constrained systems is of prime importance, the
   Hamilton-Jacobi theory provides a bridge between classical and
   quantum mechanics.
   The first study of the Hamilton-Jacobi equations for arbitrary first-order
   actions was initiated by Santilli~\cite{santilli}. Later on the canonical
   method has been developed~\cite{eqab}. In this method the equations of
   motion are written as total differential equations, and the set of
   HJPDEs are obtained.

 In an earlier work, quantum electrodynamics theory and pure
 Yang-Mills are treated as constrained systems using the canonical
 approach ~\cite{salam}. Later on this work has been extended to
 quantum chromodynamics theory, which is a Yang-Mills theory in the
 presence of quark fields \cite{hatem}. In this paper we treat one
 of the extended objects as a constrained system which is the 'tHooft-Polyaov
 monopole. In section 2, the canonical formulation is briefly reviewed and in
 section 3 a brief discussion of 'tHooft-Polyakov monopole is given. While
 in section 4 the 'tHooft-Polyakov monopole is treated as a constrained
 system and the equations of motion for the system are obtained.
 Also the quantization of the system is discussed.
%
%
\section*{2 Canonical formulation}

In this formulation, if we start with a singular Lagrangian with Hessian
matrix of rank $(N-r), r < N$, then the set of HJPDEs is expressed as
\begin{equation}
H^\prime_\alpha \left(x_\beta,q_a,\frac{\partial S}{\partial q_a},
   \frac{\partial S}{\partial x_\beta}\right) =0, \,\,\,\,\,\,
\alpha,\beta=0,1,\cdots,r \label{hjpds}
\end{equation}
where
\begin{eqnarray}
H^\prime_0 & =& p_0+H_0 , \nonumber \\
H^\prime_\mu & = & p_\mu + H_\mu , \,\,\,\,\,\,\
\mu=1,\cdots,r \label{setofhjpde}
\end{eqnarray}
and $S$ is the total action of the theory under consideration.

According to ref.~\cite{eqab,eqab1}, the equations of motion are written as
total differential equations as follows,
\begin{eqnarray}
dq_a & = &\frac{\partial H^\prime_\alpha}{\partial p_a} dx_\alpha ,
\,\,\,\,\,\, a=1,2,\cdots,N-r     \label{eofm1} \\
dp_i & = & -\frac{\partial H^\prime_\alpha}{\partial q_i} dx_\alpha,
\,\,\,\,\,\, i=1,2,\cdots,N.    \label{eofm2}
\end{eqnarray}
Following Rabei~\cite{eqab1}, these equations are integrable if
and only if
\begin{equation}
dH^\prime_\mu=0. \label{dhmu}
\end{equation}
The canonical action $S$ can easily be obtained from the following
equation~\cite{eqab,eqab1}
\begin{equation}
dS[q_i;t]= \left(-{\mathcal{H}}_{\alpha}
  + p_a \frac{\partial{\mathcal{H}}^{\prime}_{\alpha}}{\partial p_a}\right)
    dx_{\alpha}. \label{action}
\end{equation}
Note that the canonical action integral is obtained in terms of
the canonical coordinates. also, if condition (\ref{dhmu}) is not
satisfied, one consider it as new constraints and again tests the
consistency condition. Hence, the canonical formulation leads to
obtain the set of canonical phase space coordinates $q_a$ and
$p_a$ as functions of $x_\alpha$. The Hamiltonian $H'_\alpha$ are
considered as the infinitesimal generators of canonical
transformations given by parameters $x_\alpha$ respectively. In
this case, the path integral representation may be written as
\cite{muslih}
\begin{eqnarray}
<{\rm Out}|S|{\rm In}>=\int \prod_{a=1}^{N-r} dq_a dp_a {\rm exp}
\left[ \, i \int_{x_\alpha}^{x'_\alpha}
\left(-{\mathcal{H}}_{\alpha}
  + p_a \frac{\partial{\mathcal{H}}^{\prime}_{\alpha}}{\partial p_a}\right)
    dx_{\alpha}\right]. \label{quantize}
\end{eqnarray}

\section*{3 'tHooft-Polyakov monopole}
Perturbation theory is based upon making power expansions of the
path integral around the trivial vacua \cite{peskin}. However,
there are solutions of the classical, nonlinear equations of
motion that exhibit particle-like behavior that gives powerful
insight into the nonperturbative behavior of these theories. A new
quantum expansion can be developed around each exact solution,
allowing us to explore regions that are not accessible by standard
perturbation theory. In particular, these solutions give us
nonperturbative information about important physical phenomena
such as tunnelling and bound states.

In this paper we consider one type of classical solutions which is
the monopole, or particles with magnetic charge, where first
investigated by Dirac \cite{Dirac}, then lately by 'tHooft
\cite{hooft} and Polyakov \cite{polyakov}. They have been found in
gauge theory with spontaneous symmetry breaking and may have
cosmological significance \cite{vilenkin}.

The model consists of Higgs scalar fields $\phi_a(\vec x,t)$ and
vector fields $W^\mu_a,(\vec x,t)$ in ($3+1$) dimensions. The
index $a=1,2,3$, is an internal space index, which will transform
according to local (space-time dependent) $SO(3)$ transformation
given below. For any given $a$, $\phi_a$ is a scalar and
$W^\mu_a(\mu=0,1,2,3)$ is a vector under Lorentz transformation.
 The Lagrangian density is
\begin{equation}
{\mathcal L}=-\frac{1}{4} G_a^{\mu\nu} G_{a\mu\nu}+\frac{1}{2}
D^{\mu}\vec\phi {\cdot} D_\mu \vec\phi - V(\vec\phi),
\label{lagrang}
\end{equation}
where $\vec\phi$ is the  Higgs field and is given by
$\vec\phi=(\phi_1,\phi_2,\phi_3)$, and the potential is given by
\begin{equation}
V(\vec\phi)=\frac{1}{4} \lambda
(\phi_1^2+\phi_2^2+\phi_3^2-\sigma^2)^2, \label{potentail}
\end{equation}
where $\phi^2=\sigma^2$ is the Higgs vacuum. The gauge field
strength is $G_a^{\mu\nu}$ and given by
\begin{equation}
G_a^{\mu\nu}=\partial^\mu W_a^\nu - \partial^\nu W_a^\mu-e
\epsilon_{abc} W_b^\mu W_c^\nu.
\end{equation}
The Lagrangian density is invariant under the group $SO(3)$ which
is generated by $T^a$ such that
\begin{equation}
U=e^{i \theta^a T^a/\sigma},
\end{equation}
where $\theta^a$ is the group parameters vary with space-time.

 Let the monopole configuration be centered at the origin.
Energy finiteness implies that there is some radius $r_0$ such
that for $r\geq r_0$
\begin{equation}
D^{\mu} \vec\phi=\partial^\mu\vec\phi-e{\vec W}^\mu \times
\vec\phi=0, \label{dmu}
\end{equation}
and
\begin{equation}
\phi_1^2+\phi_2^2+\phi_3^2-\sigma^2=0, (\Rightarrow V(\vec\phi)=0)
\label{vacuum}.
\end{equation}
Regions of space-time, where the eqns.~(\ref{dmu}) and
~(\ref{vacuum}) are satisfied constitute the Higgs vacuum.

The general form of $\vec W^\mu_a$ satisfying eqn.~(\ref{dmu}),
provided $\vec\phi$ satisfies eqn.~(\ref{vacuum}) is
\begin{equation}
\vec W^\mu=\frac{1}{\sigma^2 e} \vec\phi \times \partial^\mu
\vec\phi +\frac{1}{\sigma}\vec\phi A^\mu,
\end{equation}
where $A^\mu$ is an arbitrary gauge field. So, we can write the
gauge field strength as
\begin{eqnarray}
\vec G^{\mu\nu} & = & \partial^\mu (\frac{1}{\sigma^2 e}\vec\phi
\times
\partial^\nu\vec\phi+\frac{1}{\sigma} \vec\phi A^\nu) -
\partial^\nu (\frac{1}{\sigma^2 e}\vec\phi \times
\partial^\mu\vec\phi+\frac{1}{\sigma} \vec\phi A^\mu) \nonumber \\ & - &
e (\frac{1}{\sigma^2 e}\vec\phi \times \partial^\mu \vec\phi +
\frac{1}{\sigma} \vec\phi A^\mu)\times (\frac{1}{\sigma^2
e}\vec\phi \times
\partial^\nu \vec\phi + \frac{1}{\sigma}
\vec\phi A^\nu) \nonumber \\
& = & \frac{1}{\sigma}~ \vec\phi F^{\mu\nu}
\end{eqnarray}
where
\begin{equation}
F^{\mu\nu}=\frac{1}{\sigma^3 e} \vec\phi \cdot (\partial^\mu
\vec\phi \times
\partial^\nu \vec\phi) + \partial^\mu A^\nu - \partial^\nu A^\mu.
\end{equation}
In the region outside the localized region, where
eqns.~(\ref{dmu}) and (\ref{vacuum}) are satisfied, i.e. in the
Higgs vacuum, $\cal L$ is given by
\begin{eqnarray}
{\cal L} & = & - \frac{1}{4} F^{\mu\nu} F_{\mu\nu} \nonumber \\
& = & - \frac{1}{4} [ \frac{1}{\sigma^6 e^2} \epsilon_{ijk}
\epsilon_{rst} \phi_i \phi_r \partial^\mu \phi_j \partial^\nu
\phi_k \partial_\mu \phi_s \partial_\nu \phi_t   + 2(\partial^\mu
A^\nu - \partial^\nu A^\mu) \partial_\mu A_\nu
\nonumber \\
& + & \frac{4}{\sigma^3 e} \epsilon_{ijk} \phi_i
\partial^\mu\phi_j \partial^\nu \phi_k \partial_\mu A_\nu ].
\label{lagrange}
\end{eqnarray}
\section*{4 The monopole as a constrained system}
 The coordinates are $\phi_l(\vec x)$ and $A_\mu(x)$ and their corresponding
conjugate momenta are $\pi_l(\vec x)$ and $\Pi_\mu(\vec x)$
respectively. They are given by the following expressions:
\begin{eqnarray}
\pi_l(\vec x) & = & \frac{{\partial \cal L}}{\partial(\partial_0
\phi_l(\vec
x))} \nonumber \\
 & = & \frac{\epsilon_{ijl}}{\sigma^3 e} \phi_i \partial^k\phi_j
 \left(\frac{\epsilon_{rst}}{\sigma^3 e} \phi_r \partial_0\phi_s
 \partial_k\phi_t+\partial_0 A_k - \partial_k A_0  \right ).
 \label{pil} \\
\Pi_\mu(\vec x) & = & \frac{{\partial \cal
L}}{\partial(\partial_0 A_\mu(\vec x))} \nonumber \\
& = & \frac{\epsilon_{rst}}{\sigma^3 e} \phi_r \partial_\mu\phi_s
\partial_0\phi_t +\partial_\mu A_0-\partial_0 A_\mu.
\end{eqnarray}
The nonvanishing Poisson brackets are
\begin{eqnarray}
\{\phi_i(x),\pi_j(x')\}& =& \delta_{ij}\, \delta^{(3)}(x-x'). \\
\{A^\mu(x),\Pi_\nu(x')\}& =& \delta_\nu^\mu \, \delta^{(3)}(x-x').
\end{eqnarray}
Upon quantization, these brackets have to be converted into proper
commutators.
The spatial components of the conjugate momentum of $A_\mu$ field
read as
\begin{eqnarray}
\Pi_i &= & \frac{\epsilon_{rst}}{\sigma^3 e} \phi_r
\partial_i\phi_s
\partial_0\phi_t +\partial_i A_0-\partial_0 A_i \nonumber \\
& = & F_{i0}, \label{pii}
\end{eqnarray}
where $i=1,2,3$ and the time component is
\begin{equation}
\Pi_0=0 \label{pi}={\cal H}_1,
\end{equation}
which is a constraint.

Note that using eqn.~(\ref{pii}), one can write the velocity
$\partial_0 A$ in terms of the momenta $\Pi_i$ as
\begin{equation}
\partial_0 A_i=\frac{\epsilon_{rst}}{a^3 e} \phi_r \partial_i
\phi_s \partial_0 \phi_t+\partial_i A_0 -\Pi_i. \label{partialA}
\end{equation}
Substituting eqn.~(\ref{partialA}) into eqn.~(\ref{pil}), we get
\begin{eqnarray}
\pi_l& =& - \frac{\epsilon_{lrt}}{\sigma^3 e} \phi_r
\partial^k\phi_t \Pi_k \nonumber \\
&=& -{\mathcal H}_{2l}, \label{h2l}
\end{eqnarray}
which is also a constraint. Note that eqns.~(\ref{pi}) and
(\ref{h2l}) are called primary constraints according to Dirac.

The Hamiltonian density can be obtained as
\begin{eqnarray}
{\cal H}_0 & = & \partial_0A_i \Pi^i-\partial_0\phi_l {\mathcal
H}_{2l} -
{\cal L} \nonumber \\
& =& -\frac{1}{2} \Pi_i \Pi^i -\partial_iA_0\Pi^i +\frac{1}{4}
F_{ij} F^{ij}.
\end{eqnarray}
Thus the total Hamiltonian is given by
\begin{equation}
H_0=\int(-\frac{1}{2} \Pi_i \Pi^i -\partial_iA_0\Pi^i +\frac{1}{4}
F_{ij} F^{ij}) d^3x.
\end{equation}
Using eqn.~(\ref{setofhjpde}), the set of HJPDEs reads as
\begin{eqnarray}
{\cal H}_0^\prime &=& P_0 + {\cal H}_0 , \\
{\cal H}_1^\prime &=& \Pi_0=0 , \label{hjpde1} \\
{\cal H}_{2l}^\prime &=& \pi_l+ {\cal H}_{2l}= \pi_l +
\frac{\epsilon_{lrt}}{\sigma^3 e} \phi_r
\partial^k\phi_t \Pi_k,\label{hjpde2}
\end{eqnarray}
where $P_0=\frac{\partial S}{\partial t}$,  $\pi_l=\frac{\partial
S}{\partial \phi_l}$, $\Pi_\mu=\frac{\partial S}{\partial A_\mu}$,
where $S=S[A_\mu,\phi_l;t]$ represents the action.

The above equations can be written in a more compact form as
\begin{eqnarray}
\frac{\partial S}{\partial t}+{\cal H}_0&=&0,  \nonumber \\
\frac{\partial S}{\partial A_0}&=& 0, \\
\frac{\partial S}{\partial \phi_l}+{\cal H}_{2l}&=&0,
\end{eqnarray}
and their simultaneous solutions determine the action $S$.

Using eqns.~(\ref{eofm1}) and (\ref{eofm2}), the total
differential equations are
\begin{eqnarray}
dA_\mu& = & \frac{\partial{\cal H}_0^\prime}{\partial\Pi_\mu} dt +
\frac{\partial{\cal H}_1^\prime}{\partial\Pi_\mu} dA_0 +
\frac{\partial{\cal H}_{2l}^\prime}{\partial\Pi_\mu} d\phi_l
 \nonumber \\
dA_i&=&(-\Pi_i+\partial_iA_0)dt , \\
d\Pi_\mu & = & -\frac{\partial{\cal H}_0^\prime}{\partial A_\mu}
dt - \frac{\partial{\cal H}_1^\prime}{\partial A_\mu} dA_0 -
\frac{\partial{\cal H}_{2l}^\prime}{\partial A_\mu} d\phi_l
\nonumber \\
d\Pi_i&=&-\partial_lF^{li}dt , \label{dedpi1} \\
d\Pi_0&=& \partial^i\Pi_i dt \label{dedpi0}, \\
d\pi_l& = & -\frac{\partial{\cal H}_0^\prime}{\partial\phi_l} dt -
\frac{\partial{\cal H}_1^\prime}{\partial\phi_l} dA_0 -
\frac{\partial{\cal H}_{2t}^\prime}{\partial\phi_l} d\phi_t
 \nonumber \\
 &=& -\frac{1}{2}\frac{\epsilon_{lrt}}{\sigma^3 e}
 \partial_i\phi_r \partial_j\phi_t F^{ij} dt+
 \frac{\epsilon_{lrt}}{\sigma^3 e} \phi_r \partial_i F^{i0}
 d\phi_t. \label{dedpi2}
\end{eqnarray}
These equations are integrable iff the total differential of
eqns.~(\ref{hjpde1}) and (\ref{hjpde2}) are equal to zero.

The vanishing of the total differential of ${\cal H}_1^\prime$
leads to a new constraint
\begin{equation}
{\cal H}_1^{\prime\prime}=d{\cal H}_1^\prime=
d\Pi_0=\partial^i\Pi_i dt=0.
\end{equation}
Taking the total derivative of the above equation gives
\begin{eqnarray}
d{\cal H}_1^{\prime\prime}&= & \partial^i d\Pi_i dt \nonumber \\
&=& -\partial_i \partial_j F^{ij} dt = 0,
\end{eqnarray}
which can be shown equivalent to
\begin{eqnarray}
d{\cal H}_1^{\prime\prime}&=&-\partial_\mu \partial_\nu F^{\mu\nu}
dt = 0, \nonumber \\
&=& \partial_\mu K^\mu dt=0,
\end{eqnarray}
where
\begin{eqnarray}
K^\mu & =&- \partial_\nu F^{\mu\nu}, \nonumber \\
&=& -\frac{1}{2\sigma^3 e}\epsilon^{\mu\nu\rho\eta} \epsilon_{lrt}
\partial_\nu\phi_l \partial_\rho\phi_r \partial_\eta\phi_t
\end{eqnarray}
is the conserved magnetic current provided $A_\mu=0$, for details
see~\cite{rajaraman}.

The first set of the Euler-Lagrange equations of motion can be
obtained using eqn.~(\ref{dedpi1}) and the constraint
eqn.~(\ref{dedpi0}),
\begin{equation}
\partial_\mu F^{\mu\nu}=0,
\end{equation}
which represents the equations of motion for the gauge field
$A_\mu$ and can be obtained if we take the variation of the
Lagrange density, eqn.~(\ref{lagrange}), with respect to the gauge
field $A_\mu$.

The second set of Euler-Lagrange equations of motion can be
obtained using the integrability condition eqn.~(\ref{hjpde2}),
\begin{eqnarray}
d{\mathcal H}_{2l}^\prime =0.
\end{eqnarray}
It gives
\begin{equation}
\frac{1}{2} \epsilon_{lrt}
\partial_\mu \phi_r
\partial_\nu \phi_t F^{\mu\nu}=0,
\end{equation}
which represents the equation of motion for the scalar field
$\phi_l$.

The action can be calculated using eqn.~(\ref{action}) as,
\begin{eqnarray}
dS[A_\mu,\phi_l;t]& =& (-{\cal H}_0^\prime +\Pi_i
\frac{\partial{\cal
H}_0^\prime}{\partial\Pi_i})dt \nonumber \\
& + & (\Pi_0+ \Pi_i \frac{\partial{\cal
H}_1^\prime}{\partial\Pi_i})dA_0 \nonumber \\
& + & (-{\cal H}_{2l}^\prime + \Pi_i \frac{\partial{\cal
H}_{2l}^\prime}{\partial\Pi_i})d\phi_l.
\end{eqnarray}
Hence,
\begin{equation}
S[A_\mu,\phi_l;t]=\int d^{4}x (-\frac{1}{2} \Pi_i \Pi^i
-\frac{1}{4} F_{ij} F^{ij}).
\end{equation}
We see that the original Lagrangian eqn.~(\ref{lagrange}) can be
recovered using the definition of the canonical momenta.

Note that although $\phi_l$ is introduced as coordinates in the
Lagrangian, the presence of the constraints and the integrability
conditions force us to treat it as a parameter like $t$. Since the
set of total differential equations is integrable, the canonical
phase space coordinates $A_i$ and $\Pi_i$ are obtained in terms of
independent parameters $t$ and $\phi_l$. Hence, using
eqn.~(\ref{quantize}), the path integral representation for the
system is calculated as
\begin{eqnarray}
<{\rm Out}|S|{\rm In}>=\int \prod_i dA_i d\Pi_i {\rm exp} \left[
\, i \int \left(-\frac{1}{2} \Pi_i \Pi^i -\frac{1}{4} F_{ij}
F^{ij}\right)
    d^4x\right]. \label{path}
\end{eqnarray}
One should notice that the path integral eqn.~(\ref{path}) has no
singular nature and it is an integration over the canonical phase
space coordinates $A^i$ and $\Pi^i$. Moreover, it is no need to
choose a gauge fixing and ghost fields, while in the usual path
quantization one must choose a gauge fixing and introduce ghost
fields \cite{faddeev}.
\section*{5 Conclusion}
The 'tHooft-Polyakav monopole is treated as a constrained system
using the canonical method. It is observed that the Hamilton
equations of motion are obtained to be in exact agreement with the
Euler-Lagrange equations. In Addition, the conserved current has
been obtained from the integrability conditions. In our approach,
we recover action of the 'tHooft-Polyakav monopole from the
equations of motion as well as from the integrability conditions
without redefining the Lagrange multipliers which are necessary in
Dirac's approach~\cite{govaerts}. Finally, the path integral
quantization is obtained directly as an integration over the
canonical phase space coordinates without choosing an appropriate
gauge fixing.
%
\section*{Acknowledgment}
The author would like to thank Eqab Rabei for suggestion this work
as well as for the insightful comments on the manuscript.

\end{document}